\documentclass[twocolumn,showpacs,preprintnumbers,superscriptaddress,amsmath,amssymb]{revtex4-1}
\usepackage{graphicx}
\usepackage{dcolumn}
\usepackage{bm}

\begin{document}

\title{Quantum Waveguide Theory of the Josephson effect in Multiband Superconductors }
\author{C. Nappi}
\email{ciro.nappi@spin.cnr.it}
\affiliation{%
Consiglio Nazionale delle Ricerche, Istituto superconduttori, materiali innovativi e dispositivi (CNR-SPIN),
Sezione di Napoli,  I-80078 Pozzuoli, Napoli (NA), Italy}%

\author{F. Romeo}%
\email{fromeo@sa.infn.it}
\affiliation{%
Dipartimento di Fisica "E. R. Caianiello", Universit\`{a} di Salerno,\\ I-84084 Fisciano,  Salerno (SA), Italy}%

\author{E. Sarnelli}
\affiliation{%
Consiglio Nazionale delle Ricerche, Istituto superconduttori, materiali innovativi e dispositivi (CNR-SPIN),
 Sezione di Napoli,  I-80078 Pozzuoli, Napoli (NA), Italy}%
\author{R. Citro}
\affiliation{%
Dipartimento di Fisica "E. R. Caianiello", Universit\`{a} di Salerno,\\ I-84084 Fisciano,  Salerno (SA), Italy}%
\affiliation{%
Consiglio Nazionale delle Ricerche, Istituto superconduttori, materiali innovativi e dispositivi (CNR-SPIN),
Sezione di Salerno,  I-8084 Fisciano,  Salerno (SA), Italy}%

\date{\today}

\begin{abstract}
We formulate a quantum waveguide theory of the Josephson effect in multiband superconductors, with special
emphasis on iron-based materials. By generalizing the boundary conditions of the scattering problem, we first
determine the Andreev levels spectrum and then derive an explicit expression for the Josephson current which
generalizes the formula of the single band case. In deriving the results, we provide a second quantization
field theory allowing to evaluate the current-phase relation and the Josephson current fluctuations
in multiband systems. We present results for two different order parameter symmetries,
 namely $s_\pm$ and $s_{++}$, which are relevant in multiband systems. The obtained
  results show that the $s_\pm$ symmetry can support $\pi$ states which are absent in the $s_{++}$ case.
   We also argue a certain fragility of the Josephson current against phase fluctuations in the $s_{++}$ case.
    The temperature dependence of the Josephson critical current is also analyzed and we find, for both the order
     parameter symmetries, remarkable violations of the Ambegaokar-Baratoff relation. The results are relevant
      in view of possible experiments aimed at investigating the order parameter symmetry of multiband
      superconductors using mesoscopic Josephson junctions.
\end{abstract}

\pacs{74.45.+c, 74.50.+r, 74.70.Xa, 85.25.Cp}

\maketitle

\section{Introduction}
Multiband superconductivity \cite{multiband-superc1959} has been theoretically suggested few years later the BCS formulation of the
superconducting state \cite{bcs}. The discovery of the superconductivity in MgB$_{2}$ \cite{mgb2} in 2001 and the observation of
 superconductivity in Fe-based compounds (FeBS) \cite{kami,stew} in 2008, renewed the interest towards multiband superconductivity.
  One main reason is the comprehension of the pairing symmetry in this class of superconductors which is still object of intensive
   investigation.
In the case of MgB$_{2}$ there is a clear consensus towards the picture of two coexisting in-phase superconducting gaps ($s_{++}$
pairing) \cite{giubileo2001}, while in FeBS the experimental evidences seem to be favorable, in some cases, to the $s_{\pm}$ pairing,
implying that both the electron-like and the hole-like band develop an \textit{s}-wave superconducting state with order parameters of
 opposite sign \cite{s_pm_symm}. Thus, in the $s_{\pm}$ symmetry, the superconducting gap exhibits a sign reversal between $\alpha$
 and $\beta$ bands which is absent in the $s_{++}$ symmetry. This phase difference can be probed using point contact Andreev
 reflection spectroscopy (PCARS) which is considered as one of the high-resolution phase-sensitive techniques to investigate the
 superconducting order parameter \cite{gonn,dagh,park1,kash}. This technique has recently been applied to gain insight on the
  properties of the FeBS \cite{chen,lu}. However, in contrast to the high-$T_c$ cuprate superconductivity manifesting a $d$-wave
   symmetry, phase sensitive experiments appear difficult in FeBS since the $s_{++}$ and $s_{\pm}$ pairings have the same
   crystallographic symmetry and, as a matter of fact, so far no experiment has been decisive in discriminating between the
   two symmetries. A complementary tool to gain information on the symmetry of the order parameter is the current-phase relation
    of the Josephson current. In fact, along with PCARS, the Josephson effect has been used in the past as a probe of electronic
     properties in superconductors, included the order parameter symmetry (e.g.\cite{ili,testa1,testa2,d0d0}). \\
In this paper, in order to discriminate between different symmetries of the order parameter ($s_{++}$ and $s_{\pm}$), we develop
a quantum waveguides approach for the Josephson effect in a multiband superconducting junction. The presence of more than one
band in the superconductor implies that extra scattering channels are present at the interface, a physical situation which
is analogous to a quantum waveguides theory problem \cite{Araujo,Xia} as recently suggested by PCARS modeling of a
normal-multiband superconductor junction \cite{Romeo}. We evaluate the
Josephson current carried by Andreev bound levels and demonstrate that several distinctive features of
 the $s_{\pm}$ and $s_{++}$ symmetries of the order parameter can be evidenced in the current-phase relation.
 The results are particular relevant with respect to the fast progresses in nanofabrication techniques
 \cite{wu} which allow now to explore the Josephson effect in the mesoscopic junction regime, where
  supercurrent flows through a small number of channels. In this kind of nanometric junctions the
  effect of a different symmetry, $s_{\pm}$ or $s_{++}$, should emerge and the experimental results
   easily compared with the basic theory of the Josephson effect in these materials.\\
The paper is organized as follows. In Section \ref{sec:model} we present the quantum waveguide model
 of a multiband superconductor coupled to another identical multiband superconductor (symmetrical
  junction) and derive the spectral equation of the Andreev bound states. In Section
  \ref{sec: josephson} we calculate the Josephson current by using a field theory formalism
  in second quantization.  In Section \ref{sec:results} we discuss the results for the
   current-phase relation and the temperature dependence of the critical current.
    In Section \ref{sec:conclusions} we draw the conclusions.
     Details on the computation of the scattering coefficients are
      reported in Appendix \ref{appendix:matrix}. The theory
      of the magnetic field dependence of the critical current
      of a Josephson junction is briefly recalled in Appendix \ref{appendix:Ic-mfield}.

\begin{figure}[htbp]
\centering
\includegraphics[width=15pc]{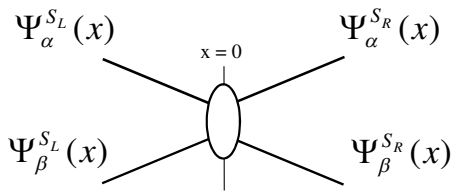}
 \caption{Quantum waveguide schematization of a two-band Josephson
 junction. Each branch represents a superconducting band, $\alpha$
 or $\beta$, in the left ($S_L$) or right ($S_R$) electrode.
 The wave functions of supeconducting bands are given by
 $\Psi^{S_{L,R}}_{\alpha, \beta}(x)$. The node at $x=0$
 represents the interface between the left and right
 superconducting banks.}\label{fig:waveguide}
\end{figure}

\section{Model and Theory }
\label{sec:model}
We formulate a ballistic theory of the Josephson
effect describing multiband superconducting junctions.
 The theory allows to consider an arbitrary number of
 bands, which are treated as network branches of an effective
 quantum waveguide model. The proposed approach could be
  applied to the $\mathrm{Mg}\mathrm{B}_2$ as well as the
   FeBS case. In order to develop the theory, hereafter we refer
   to the specific case of FeBS for which the $s_{++}$ and $s_{\pm}$ symmetries
    have been suggested. The Josephson effect in FeBS has been already studied
    by using a number of different methods \cite{burm2,burm1,golu,nap,moor,
     da, stan, lin, apos, kosh, berg, ota, erin, park, wu1, ng, golu2, lind, tsai, chen2}.
      However, most of these studies, except few cases \cite{nap,moor,da}, deal with the
       so-called hybrid Josephson junction in which the junction is formed by
        a conventional (s-wave) superconductor and a ($s_{\pm}$ or $s_{++}$) FeBS.
         We specialize here to the case of an all-FeBS coplanar Josephson junction,
          in which both the electrodes are FeBS materials (symmetric junction),
          as occurs for instance in grain boundary junctions. So far, several Josephson
          junctions using thin films have been fabricated
          \cite{seba, dor,dor2,schm, lee, kata1, kata2, sarn, wu2,baro}
           on bicrystal substrates (for a review on Fe-based Josephson junctions,
            see Ref.\cite{seidel} and references therein) and thus a theoretical
             effort along this direction is needed.\\
Thus we consider a Josephson junction model in which the
 electrodes are two-band FeBS. To this end the junction
 is represented by a
network of four one-dimensional branches connected to a
single node point $x=0$ (see Fig.~\ref{fig:waveguide}),
 $x$ being the coordinate along the propagation direction
  normal to the interface. Each superconducting branch
  represents the $i$-th band on the left or the right
  side of the junction, while the band wave function
  $\Psi_i=\theta(-x)\Psi^{S_{L}}_i(x)+\theta(x)\Psi^{S_{R}}_i(x)$,
   in the clean limit, obeys the equation
\begin{eqnarray}
\label{eq:hamilton}
\left[\begin{array}{rl}
&\hat{H}_j(x)  \hspace{0.5 cm} \Delta_j(x)   \\
&\Delta^*_j(x)  \hspace{0.5 cm}  -\hat{H}^*_j(x)
\end{array}\right]\Psi_j=E \Psi_j
\end{eqnarray}
where $j \in \{\alpha,\beta \}$ is the band index, while
 \begin{equation}
\label{eq:pair_potential}
 \Delta_j(x)=\Delta_je^{i \varphi_j}\theta(-x)+\Delta_je^{i
(\varphi_j+\varphi)}\theta(x),
\end{equation}
($\theta(x)=0$, for $x \leq 0$, $\theta(x)=1$ for $x>0$) are
the two coexisting pair potentials. The operator $\hat{H}_j(x)$
represents the single particle Bogoliubov-de Gennes Hamiltonian
 in the two bands, which reads
\begin{equation}
\label{eq:hamilton2}
\hat{H}_j(x)=-\frac{\hbar^{2}}{2 m_j}\frac{\partial^2}{\partial x^2}-E_F.
\end{equation}
In writing Eq.~(\ref{eq:pair_potential}) we are neglecting the proximity
effect and assuming that the inhomogeneous character of the gap
in the junction is captured by choosing $\Delta_\alpha=\Delta_\beta=0$
just at the node (short junction). The two gap values
 $\Delta_\alpha$, $\Delta_\beta$, with $\Delta_\alpha < \Delta_\beta$,
 are assumed to be the same in the
two superconductive leads. The quantity $\varphi$ is the gauge
 invariant phase difference between the two superconductive regions,
  $\varphi_\alpha$ and  $\varphi_\beta$ the internal pair potential
  phases. In the case of $s_{\pm}$-wave gap model,
   $\varphi_\beta-\varphi_\alpha=\pi$ and the two gap
   have opposite sign, while in the standard two band
   model, with same sign gaps, $\varphi_\beta-\varphi_\alpha=0$.
 The quantities $m_j$ in Eq.~(\ref{eq:hamilton2}) are
 the effective masses of quasiparticles in the $j$-th
 superconducting  branch and are material-dependent quantities.
We also introduce a single-particle node potential $U(x)$
 which is different from zero only for $x=0$ and can be
 modeled as the usual Blonder-Tinkham-Klapwijk \cite{blond}
  interface potential $U(x)=U_{0}\delta(x)$, even though
  this limiting assumption is not required to develop the
  theory. The node potential $U(x)$ allows the modeling of
  a FeBS/insulator/FeBS or FeBS/normal-metal/FeBS junction
  in which a normal scattering at the interface reduces the
   transparency of the junction. The junction barrier
    strength can be still characterized by introducing a
    Blonder-Tinkham-Klapwijk dimensionless parameter
    $Z=mU_0/(\hbar^2 k_F)$ ($m$ being the bare electron mass
     and $k_F^{2}=2 m E_F/\hbar^2$) within the boundary
     conditions of the scattering problem. The modified
     boundary conditions used here (see Eqs.(\ref{eq:matching1}),
      (\ref{eq:matching2})) account for band-sensitive
      scattering effects and have been already introduced
      in Ref.~\cite{Romeo} to describe the differential
      conductance of a normal-metal/FeBS junction; in the
       following we discuss their generalization to the
        Josephson junction case. The potentials $U(x)$
        and $\Delta_j(x)$ are responsible for the normal
         scattering and the scattering of
electrons into holes (Andreev scattering) at the interface,
 respectively. The four wave functions,
  $\Psi^{S_{L}}_j(x)$, $\Psi^{S_{R}}_j(x)$ ($j=\alpha,\beta$),
  one for each branch in the two electrodes, can be written
  in terms of the eigenstates of the local Hamiltonians as
\begin{eqnarray}
\label{eq:PSISd}
&&\Psi^{S_{L}}_j(x) =
\left[\begin{array}{rl}
&u^L_j(x) \\
 &v^L_j(x)
\end{array}\right]=  \\
&& {a_j}\left(
\begin{array}{rl}
&v_j  \\
 &u_j e^{-i \varphi_j}
\end{array}
 \right)e^{ i p^j_h x}+  b_j\left(
\begin{array}{rl}
&u_j  \nonumber\\
 &v_j e^{-i \varphi_j}
\end{array}
 \right)e^{-i p^j_e x},
 \end{eqnarray}
for $x<0$ or
 \begin{eqnarray}
 \label{eq:sol2}
&&  \Psi^{S_{R}}_j(x) = \left[\begin{array}{rl}
&u^R_j(x)  \\
&v^R_j(x)
\end{array}\right]= \\
&&c_j\left(
\begin{array}{rl}
&u_j  \\
&v_j e^{-i( \varphi_j+\varphi)}
\end{array}
 \right)e^{ i p^j_e x}+\ d_j\left(
\begin{array}{rl}
&v_j\nonumber \\
 &u_j e^{-i (\varphi_j+\varphi)}
\end{array}
  \right)e^{-i p^j_h x},
\end{eqnarray}
for $x>0$.
Here $u_{\alpha}$ and $v_{\alpha}$, and $u_{\beta}$ and $v_{\beta}$ are
the Bogoliubov coefficients for the first and second band, respectively
\begin{eqnarray}
&&u_j=\left[\frac{1}{2}\left(1+i\frac{ \Omega_j}{E}\right)\right]^{1/2},
v_j=\left[\frac{1}{2}\left(1-i\frac{ \Omega_j}{E}\right)\right]^{1/2} \nonumber \\
\end{eqnarray}
with $\Omega_{j}=\sqrt{\Delta_{j}(T)^2-E^2}$, $j \in\{\alpha, \beta\}$, while
 $T$ represents the temperature of the thermal bath. The wave vectors of the two bands
 \begin{eqnarray}
 \label{eq:momentum}
{ p^{j}_{e,h}}=\sqrt{r_j^2 k_F^2 \pm  2im_j \Omega_j/\hbar^2}  \nonumber
 \end{eqnarray}

 are well approximated by the expressions:
\begin{eqnarray}
\label{eq:expansion}
p^\alpha _{e,h}\simeq r_\alpha k_F \left(1\pm i \chi{ \sqrt{1-E^2/{\Delta_\alpha}^2}}\right),\\
 \nonumber p^\beta _{e,h}\simeq r_\beta k_F \left(1\pm i \chi{ \sqrt{{\Delta_\beta}^2/{\Delta_\alpha}^2-E^2/{\Delta_\alpha}^2}}\right),
\end{eqnarray}
which are valid under the assumptions $E_{F}\gg \Delta_{\beta}$, while the different
 signs $"+"$,  $"-"$ refer to electronlike or holelike excitations respectively.
 The coefficient $r_j^2=m_j/m$ represents the effective mass of the $j$-th band
  measured in units of the bare electron mass $m$. In writing Eqs.~(\ref{eq:expansion})
   we have introduced the dimensionless factor $\chi=\Delta_\alpha(T)/2 E_F$ or
    $\chi=\Delta_\alpha(T)/(k_F v_F \hbar)$, which represents the ratio between
     the Fermi wavelength $k_F^{-1}$ and the coherence length $\xi(T)\sim \hbar
     v_F/\Delta_\alpha(T)$. Its zero temperature value, $\chi_0$, can be also
     defined as $\chi_0=\Delta_\alpha(0)/(2 E_F)$.
Since we are interested in localized subgap states, in Eq. (\ref{eq:expansion})
 we assume $E<\Delta_\alpha(0)$, ensuring that the wave functions $\Psi^{S_{L}}_j(x)$,
  $\Psi^{S_{R}}_j(x)$ ($j=\alpha,\beta$) decay exponentially for $|x|\rightarrow\infty$. \\
The quasiparticle wave functions obey the generalized matching quantum waveguide conditions
\cite{Romeo}
\begin{widetext}
 \begin{eqnarray}
 \label{eq:matching1}
 \Psi^{S_R}_{\alpha}(0)=s_\alpha\Psi^{S_L}_{\alpha}(0),
 \qquad \Psi^{S_R}_{\beta}(0)=s_\beta\Psi^{S_L}_{\alpha}(0),
 \qquad \Psi^{S_L}_{\beta}(0) =s\Psi^{S_L}_{\alpha}(0)
 \end{eqnarray}
\begin{eqnarray}
 \label{eq:matching2}
\frac{\partial}{\partial x}\left[s_\alpha\frac{1}{r_{\alpha}^2}\Psi^{S_{R}}_{\alpha}+
 s_\beta\frac{1}{r_{\beta}^2}\Psi^{S_{R}}_{\beta}\right]_{x=0}-
 \frac{\partial}{\partial x}\left[\frac{1}{r_{\alpha}^2}\Psi^{S_{L}}_{\alpha}+
 s\frac{1}{r_{\beta}^2}\Psi^{S_{L}}_{\beta}\right]_{x=0}=2 k_F Z  \Psi^{S_L}_{\alpha}(0).
 \end{eqnarray}
 \end{widetext}
Equations (\ref{eq:matching1}) and  (\ref{eq:matching2})
 guarantee the conservation of the charge current at the
  node $x=0$ (quantum Kirchhoff's law) and generalize the
  waveguide boundary conditions given in Ref.~\cite{Romeo}
  to the Josephson junction case. The three new parameters
  $s_\alpha,s_\beta, s$ characterize the interface and make
   the scattering and the current flow band-sensitive.
   Indeed, the overlap between the wave functions on different
   sides of the junction may favour scattering events towards
   a specific band, $\alpha$ or $\beta$. The existence of these
   'band coupling parameters' implies that a discontinuity of
   the wave functions may occur at the node ($x=0$) when at least
    one of the coupling parameters $s_\alpha,s_\beta,s$ is
    different from one. The meaning of this discontinuity
    has been discussed in Ref.~[\onlinecite{Romeo}] in
    connection with the theory of PCARS in multiband
    superconductors. The physical meaning of the overlap
    parameters $s_\alpha,s_\beta,s$ can be easily understood
     observing that setting, for instance, $s_{\alpha}=0$ implies $\Psi^{S_R}_{\alpha}(0)=0$.
      Thus the wave function $\Psi^{S_R}_{\alpha}(x)$ presents a
       vanishing overlap at the interface with the remaining branch wave functions.
        As a consequence, in this case, the band $\alpha$ on the right
        side of the junction is excluded from the transport.
        The latter point can be further clarified by looking at
        the prefactor of the first term of Eq.~(\ref{eq:matching2}),
         namely $s_{\alpha}(m_{\alpha}/m)^{-1}$, which governs the
          particle flow through the $\alpha$ band on the right side
          of the junction. Lowering $s_{\alpha}$ has the same effect
           of increasing $m_{\alpha}$, causing a lowering of the
            particles flux through the considered band.
            In the equal coupling case ($s_\alpha=s_\beta=s=1$)
            the partitioning of the current among different
             network branches is entirely governed by the bulk
              mass ratios $r_{j}$ and only weakly affected by
               the interface scattering potential strength $Z$.
               Thus the parameters
               $s_\alpha,s_\beta,s$ can be properly interpreted
               as overlap or coupling factors which define
               the fraction of current flowing through a specific band.\\
Eqs.~(\ref{eq:matching1})-(\ref{eq:matching2}) provide a
linear system of equations for the eight unknown
 coefficients $a_j$, $b_j$, $c_j$, $d_j$ ($j \in \{\alpha,\beta\}$),
  appearing in the wave functions (see Appendix \ref{appendix:matrix}).
The bound state spectrum can be determined imposing
the condition that this homogeneous system of equations
 has a nontrivial solution. Within the Andreev's
  approximation, i.e.  $p_e^{\alpha}\simeq p_h^{\alpha}\simeq r_\alpha k_F$ and
   $p_e^{\beta}\simeq p_h^{\beta}\simeq r_\beta k_F$, this critical condition is written as
\begin{widetext}
 \begin{eqnarray}
\label{eq:spectrum}
&&-\frac{s_\beta^2(\epsilon^2-\gamma^2)(1- \epsilon^2 -
2 s_\alpha^2 \epsilon^2 -s_\alpha^4(\epsilon^2-1)+
2 s_\alpha^2\cos \varphi)}{r_\alpha^2}+\nonumber \\
&&\frac{ s^2 (\epsilon^2-1)(2 s s_\beta^2 \epsilon^2+
(s^2+s_\beta^4)(\epsilon^2-\gamma^2) -2s s_\beta^2 \gamma^2 \cos \varphi)}{r_\beta^2}+ \nonumber \\
&&\frac{2 s s_\beta \sqrt{\epsilon^2-1}\sqrt{\epsilon^2-
\gamma^2}\left(s((1+s_\alpha^2)\epsilon^2-e^{i\delta}\gamma)+
s_\beta^2((1+s_\alpha^2)\epsilon^2-e^{i\delta}s_\alpha^2 \gamma)-
e^{i\delta}(s s_\alpha^2+s_\beta^2)\gamma \cos \varphi\right)}{r_\alpha r_\beta}+
\nonumber \\ &&4s_\beta^2Z^2(\epsilon^2-1)(\epsilon^2-\gamma^2)=0,
\end{eqnarray}
\end{widetext}
where $\delta=\varphi_\beta-\varphi_\alpha$, $\gamma=\Delta_\beta/\Delta_\alpha$ and $\epsilon=E/\Delta_\alpha(T)$.
Choosing $\delta=0$ or $\delta=\pi$ reflects the internal pairing symmetry $s_{++}$ or $s_{\pm}$, respectively.
\begin{figure}[h]
\centering
\includegraphics[width=20pc]{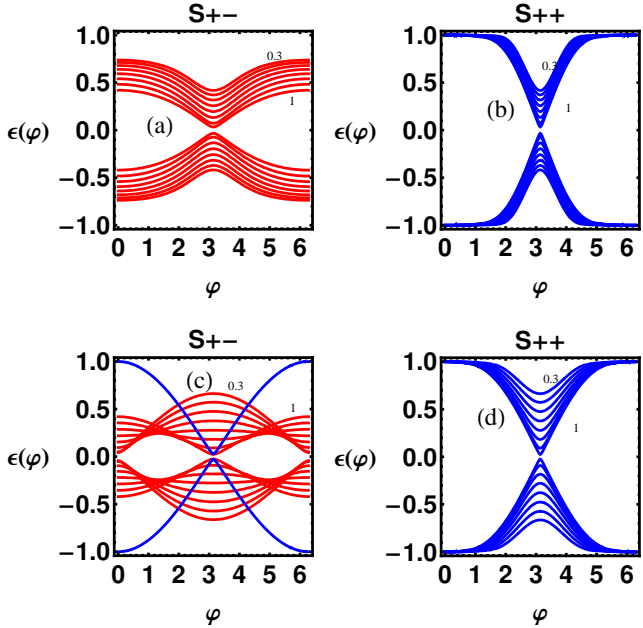}
 \caption{Bound states energy levels $E^+(\varphi)$
 and $E^-(\varphi)$ (normalized to $\Delta_\alpha(T)$)
 calculated through Eq.~(\ref{eq:spectrum}). Panels (a),
 (c): $s_{\pm}$ symmetry. Panels (b), (d): $s_{++}$ symmetry.
  $Z$ and $\gamma$ have been fixed to $Z = 0.03$ and $\gamma = 3$,
   the various curves are obtained by letting parameter $s_\alpha$
   to vary in the range $(0.3,1)$ in steps of $0.1$, while $s_\beta=s=1 $.
    Panels (a) and (b): $r_\alpha=1.8$, $r_\beta=1$. Panels (c) and (d):
     $r_\alpha=1$, $r_\beta=1.8$, the blue line curve in panel (c)
     is the Andreev level of a conventional s-wave junction with $r_jZ=0.03$. }\label{fig:levels}
\end{figure}
Two sets of energy levels $E^+(\varphi)$ and $E^-(\varphi)$ are
found in solving Eq.~(\ref{eq:spectrum}), which correspond to
electronlike and holelike quasiparticles, respectively. One finds
that the Andreev bound states never appear beyond $\Delta_\alpha$.
When $s_\alpha=s_\beta=s=1$, in the two limits $r_\alpha \rightarrow
 \infty$ or $r_\beta \rightarrow \infty$ (exclusion of the $\alpha$ or
 $\beta$ band respectively) Eq.~(\ref{eq:spectrum}) provides the well
 known s-wave result $E=E_B=\Delta_j \sqrt{(\cos^2 (\varphi/2)+Z_j^2)/(1+Z_j^2)}$,
 with $Z_{j}=r_jZ$ \cite{nota-Zj}, $j$  being $\alpha$ or $\beta$ (see Fig.~\ref{fig:levels}
  (c)). \\In Fig.~\ref{fig:levels} we show Andreev bound states spectra computed using
   Eq.~(\ref{eq:spectrum}). Once the mass ratios $r_{j}$ have been fixed, different
    Josephson couplings can be obtained depending on the kind of the order parameter
     symmetry and on the choice of the overlap factors. To better explain this point,
      we present different curves obtained by letting the parameter $s_\alpha$ to
      vary in the range $(0.3,1)$ in steps of $0.1$ while $s_\beta=s=1 $. This choice
      of the overlap factors corresponds to weaken progressively the coupling of the
       band $\Psi^{S_{R}}_{\alpha}$ from the remaining bands. Panels (a) and (b)
        ($r_\alpha=1.8$, $r_\beta=1$) show a generalized lowering of the levels
        toward the zero energy state in the case of $s_{\pm}$ symmetry (panel (a)),
         which is absent in the case of $s_{++}$ symmetry (panel (b)) and represents
          the main difference between the two symmetries in this case. A further
          difference is illustrated in panels (c) and (d)  ($r_\alpha=1$, $r_\beta=1.8$).
           In the case of $s_{\pm}$ symmetry (panel (c)) the energy levels undergo a
            gradual concavity change as the parameter $s_\alpha$ is decreased. This
            concavity change is responsible for a $\pi$-shift in the current-phase
             relation, which is not observed in the $s_{++}$ case (Fig.~\ref{fig:levels}(d)).
              Thus, as also observed elsewhere \cite{nap}, a $0-\pi$ transition may be
              a distinctive signature of the $s_{\pm}$ symmetry. This point will be
              further explained in Section \ref{sec:results}, where the current-phase
               relations and the temperature dependence of the maximum Josephson
               current will be presented.

\section{Josephson current}
\label{sec: josephson}
The Josephson current flowing through the junction  could be directly
computed from the Andreev bound states according to the formula \cite{benakk}:
\begin{equation}
\label{eq:josephson-der}
I_{J}(\varphi)\approx -\frac{2e}{\hbar}\frac{dE^{+}(\varphi)}{d \varphi}\tanh \Big[\frac{E^{+}(\varphi)}{2 k_B T}\Big].
\end{equation}
Due to the special character of the bound state of a multiband junction, in
the following we will characterize the Josephson current by using a field
theory formalism in second quantization. We verified that the latter procedure
provides results consistent with those obtained using Eq.~(\ref{eq:josephson-der}).\\
The Cooper pair charge flow $\bar{J}_{ch}$ through the junction can be computed
as quantum-statistical average of the current density operator according to the
expression $\bar{J}_{ch}=\sum_{j,\sigma}\langle\hat
{\psi}_{j\sigma}^{\dag}\check{\mathcal{J}}_{j}
\hat{\psi}_{j\sigma}\rangle$, where
 $\check{\mathcal{J}}_{j}=\frac{i \hbar |e|}{2m_{j}}
 (\overrightarrow{\partial}_x-\overleftarrow{\partial}_x)$
 represents the first-quantization band operator \cite{nota_diff_op}.
  The second quantized fields $\hat{\psi}_{j\sigma}$
  ($\hat{\psi}_{j\sigma}^{\dag}$) represent the annihilation
  (creation) operators of a spin $\sigma$ electron in the
   $j$-th band and obey a fermionic algebra. These fields
   provide an appropriate basis in the absence of superconducting
    correlations. However, when the superconductivity is established,
     $\hat{\psi}_{j\sigma}$ are not eigenfields of the problem and,
      for this reason, the Nambu representation can be conveniently
       used $\Psi(x)=\sum_{j \in\{\alpha,\beta\}}|j\rangle
        \otimes (\hat{\psi}_{j\uparrow}  , \hat{\psi}^{\dag}_{j\downarrow})^{t}$.
         The Nambu field $\Psi(x)$ is a non-local quantity
          describing particle-hole excitations in different network
           branches and can be expanded in eigenfields of the
            Hamiltonian problem on the network. Limiting
            the expansion to the eigenfields $\hat{\gamma}_{\sigma}$ describing the low-energy
             (sub-gap) states with energy $\pm E_{B} \in [-\Delta_{\alpha},\Delta_{\alpha}]$, we get:
\begin{equation}
\Psi(x) \approx \hat{\gamma}_{\uparrow}e^{-iE_{B}t/\hbar}\psi_B(x)+
\hat{\gamma}^{\dag}_{\downarrow}e^{iE_{B}t/\hbar}\tilde{\psi}_B(x),
\end{equation}
where $\psi_B(x)=\theta(-x)\psi^{(L)}_B(x)+\theta(x)\psi^{(R)}_B(x)$
 is the wave function of the electron-like bound state having
 energy eigenvalue $E_{B}>0$, while $\tilde{\psi}_B(x)=[\sum_{j}
 |j \rangle \langle j|\otimes i \hat{\sigma}_{y} \mathcal{C}]
 \psi_B(x)$ represents its time-reversed state associated with
  a hole-like state with energy $-E_{B}$. The electron-like
  bound state is localized at the interface and extends over
   all the waveguide branches. Thus eigenstates of the local
   branch Hamiltonians can be used to expand $\psi_{B}(x)$, using the following decomposition:
\begin{eqnarray}
\psi^{(\nu)}_B(x)=\sum_{j \in\{\alpha,\beta\}}|j\rangle \otimes \left[\begin{array}{rl}
&u^{\nu}_j(x)  \\
&v^{\nu}_j(x)
\end{array}\right],
\end{eqnarray}
with $\nu \in [L,R]$ an index identifying the left ($x<0$) or
right ($x>0$) side of the junction. Once $\Psi(x)$ has been
expanded in eigenfields $\hat{\gamma}_{\sigma}$ it is possible
to recognize the fermionic fields $\hat{\psi}_{j\uparrow}$ in
 the expression:
\begin{equation}
\label{eq:fieldexp}
\hat{\psi}_{j\uparrow} \approx \hat{\gamma}_{\uparrow}
e^{-iE_{B}t/\hbar}u_{j}(x)-\hat{\gamma}^{\dag}_{\downarrow}e^{iE_{B}t/\hbar}v_{j}(x)^{\ast},
\end{equation}
with $u_{j}(x)=\theta(-x)u^{L}_{j}(x)+\theta(x)u^{R}_{j}(x)$
and analogously for $v_{j}(x)$.
Substituting Equation (\ref{eq:fieldexp}) in the expression
 for $\bar{J}_{ch}$, in the absence of spin-sensitive potentials
 \cite{nota_current}, we obtain ($x>0$):
\begin{eqnarray}
\label{eq:josephson-scattering}
\bar{J}_{ch}&=&\frac{-2|e|\hbar}{m}\sum_{j}r_{j}^{-2}
\Bigl \{\mathrm{Im}[u^{R}_{j}(x)^{\ast}\partial_{x}u^{R}_{j}(x)]f(E_{B})+ \nonumber\\
&+& \mathrm{Im}[v^{R}_{j}(x)\partial_{x}v^{R}_{j}(x)^{\ast}](1-f(E_{B}))\Bigl\},
\end{eqnarray}
where we explicitly used the thermal equilibrium averages
 $\langle \hat{\gamma}^{\dag}_{\sigma}\hat{\gamma}_{\sigma}\rangle=f(E_{B})$ and
 $\langle \hat{\gamma}_{\sigma}\hat{\gamma}^{\dag}_{\sigma}\rangle=1-f(E_{B})$.
  Due to the current conservation we have the freedom to evaluate the current
   in $x=0^+$, and thus starting from Eq.~(\ref{eq:josephson-scattering})
    we get ($e=-|e|$) \cite{nota_current2}:
\begin{eqnarray}
\label{eq:current2}
\bar{J}_{ch}=-2 e v_F \sum_{j}\frac{u_j v_j}{r_j}\left( |c_j|^2-|d_j|^2\right)\tanh \left[ \frac{E_B}{2 k_B T}\right],
\end{eqnarray}
where the coefficients $c_j$ and $d_{j}$ are calculated at the energy
$E_B=\epsilon_B(\varphi) \Delta_\alpha(T)$, while $\epsilon_B(\varphi)$
is solution of Eq.~(\ref{eq:spectrum}). It's worth mentioning here that
the Fermi velocity in the $i$-th band is given by $v^{(i)}_{F}=v_F/r_{i}$,
while the quantity $v_{F}=\hbar k_F/m$ is just used as velocity unit.
Equation (\ref{eq:current2}) is one of the main results of this work
and generalizes the expression found in Ref.~[\onlinecite{furusaki}]
to the multi-band case.\\
Equation (\ref{eq:current2}), complemented by the coefficients $c_j$
 and $d_j$ derived following the procedure sketched in Appendix \ref{appendix:matrix},
  allows to obtain the current-phase relation of the junction. In deriving the
   scattering coefficients $a_j$, $b_j$, $c_j$, $d_j$ ($j\in\{\alpha,\beta\}$),
    Eqs.~(\ref{eq:matching1})-(\ref{eq:matching2}) have to be complemented by
     the normalization condition, $\int^{\infty}_{-\infty} dx \psi_{B}(x)^{\dag}\psi_{B}(x)=1$,
     of the bound state wave function
     $\psi_{B}(x)$: $\sum_{j=\alpha,\beta}\int _{-\infty}^0 [  |u^L_j(x)|^2+|v^L_j(x)|^2]dx+
      \sum_{j=\alpha,\beta}\int _0^{+\infty} [|u^R_j(x)|^2+|v^R_j(x)|^2]dx=1$.
      The normalization condition can be satisfied only considering an
       imaginary part in the expressions of the quasiparticle momenta
       $p^j_e$ and $p^j_h$. This accounts for the localized nature of
       the bound state whose wave function contains decaying
        exponentials of the type $\sim \exp[\pm \chi r_\alpha k_F \sqrt{1-(E_{B}/\Delta_{\alpha})^2} x]$
        and $\sim \exp[\pm \chi r_\beta k_F \sqrt{\gamma^2-(E_{B}/\Delta_{\alpha})^2} x]$,
         the decay length being comparable with the coherence length $\xi(T)\sim \hbar v_F/\Delta_\alpha(T)$.
          Due to the different decay lengths characterizing the bound state wave function, $|E_{B}|$
          cannot exceed $\Delta_{\alpha}$. Indeed, assuming $E_{B}>\Delta_{\alpha}$, the quantum
          state is no more localized and its wave function $\psi_{B}(x)$ is not normalizable.
          The above arguments explains why the bound state energy $E_{B}$ cannot
          exceed the minimum value among the superconducting gaps describing the
           multiband junction. This latter aspect suggests that the performances
           of a multiband Josephson junction can be strongly affected by disorder effects and inhomogeneities.

\begin{figure}[htbp]
\centering
\includegraphics[width=18pc]{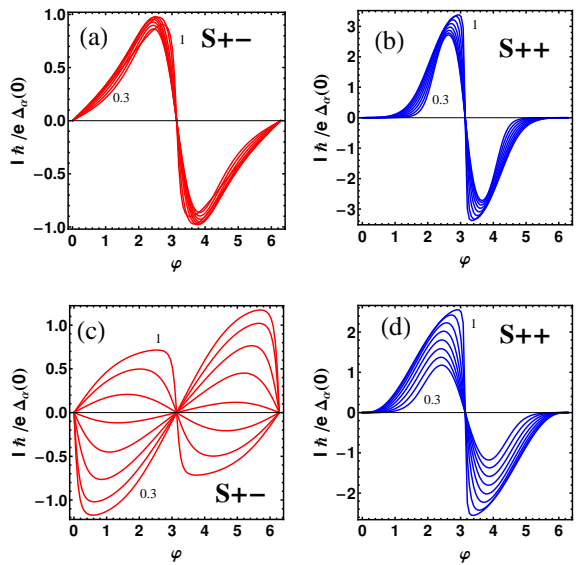}
 \caption{Current-phase relations at $T=0.001T_c$. Panels (a) and (c): $s_{\pm}$ symmetry.
 Panels (b) and (d): $s_{++}$ symmetry. Parameters are the
 same of the corresponding panels in Fig.~\ref{fig:levels}; (a)
  and (b): $Z = 0.03$, $\gamma = 3$, $r_1 = 1.8$, $r_2 = 1$; (c)
   and (d): $Z = 0.03$, $\gamma = 3$, $r_1 = 1$, $r_2 = 1.8$. The
   value of $\chi_0$ has been set to $\chi_0=0.01$. The various
   curves are obtained by letting parameter $s_\alpha$ to vary in
    the range $(0.3,1)$ in steps of $0.1$, while $s_\beta=s=1 $.}\label{fig:phase_current}
\end{figure}

\begin{figure}[htbp]
\centering
\includegraphics[width=20pc]{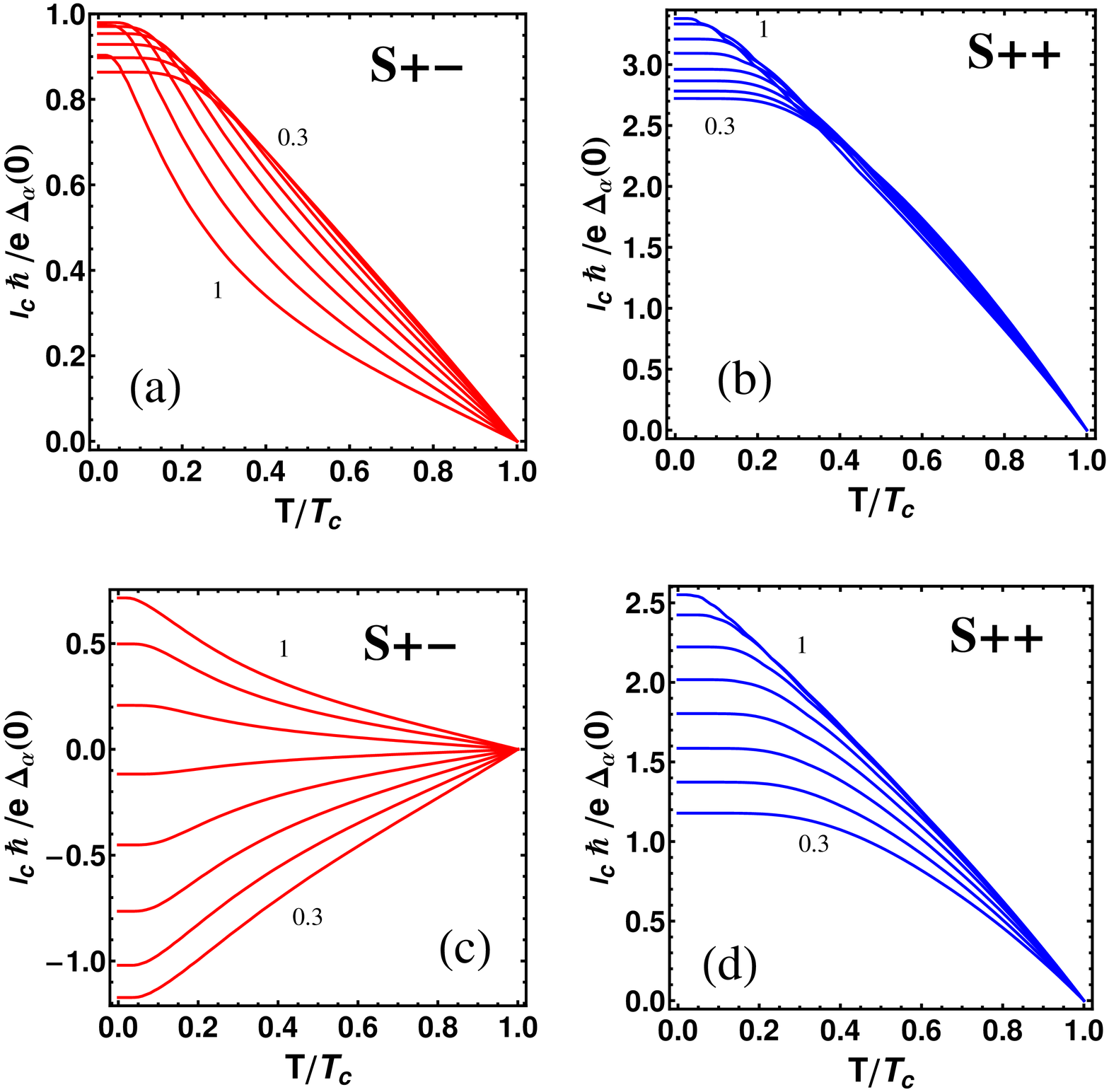}
 \caption{ Josephson critical current as a function of
  the reduced temperature $T/T_c$. Panels (a) and (c):
   $s_{\pm}$ symmetry. Panels (b) and (d): $s_{++}$ symmetry.
   Parameters are the same of the corresponding panels in
   Fig.~\ref{fig:levels}: $Z=0.01$, $\gamma = 3$.
   The value of $\chi_0$ has been set to $\chi_0=0.01$.
    (a) and (b) $r_1=1.8$, $r_2=1$. (c) and (d),
     $r_1=1$, $r_2=1.8$. The critical current is
     in units of $e\Delta_\alpha(0)/\hbar$. The
      various curves are obtained by letting
       parameter $s_\alpha$ to vary in the range
        $(0.3,1)$ in steps of $0.1$, while $s_\beta=s=1 $.}\label{fig:IcvsT}
\end{figure}
\section{Numerical Results }
\label{sec:results}
In this section we provides specific examples
of current-phase relation derived through Eq.~(\ref{eq:current2}).
 We assume that the temperature dependence of both gaps $\Delta_j$
  ($j=\alpha,\beta$) is given by $\Delta_j(T)=\Delta_j(0)\tanh\left( 1.74 \sqrt{T_c/T-1} \right)$,
   where $T_c$ is the critical temperature of the superconducting transition. We also
    assume the validity of the BCS ratio $\Delta_\alpha(0)/k_B T_c=1.76$ between the
    critical temperature and the zero temperature pair potential. Under these assumptions,
     the parameter $\gamma=\Delta_\beta/\Delta_\alpha$ does not depend on the temperature.
      Note that the quantity $\chi_0$ has been assigned the value 0.01 in the calculations
      but that the results do not depend on the specific value of $\chi_0$, provided that $\chi_0 \ll 1$.   \\
In Figure \ref{fig:phase_current} we present the current-phase relation of two junction
configurations whose bound states energy has been
shown in Fig.~\ref{fig:levels}. In particular, in panels (a) and (b), we report the current-phase
relation computed for the $s_{\pm}$ and the $s_{++}$ symmetry, respectively, setting the model parameters as follows:
$Z = 0.03$, $\gamma = 3$, $r_1 = 1.8$, $r_2 = 1$. The different curves are obtained by letting the
 parameter $s_\alpha$ to vary in the range $(0.3,1)$ in steps of $0.1$, while $s_\beta=s=1 $.
 One notices a strong suppression of the Josephson current in the $s_{\pm}$ symmetry (panel (a))
 compared to the $s_{++}$ case (panel (b)). This phenomenon originates from the destructive
  interference effects of quasiparticles experiencing two opposite gap signs. The destructive
   interference lowers the Andreev reflection probability such that, in perfectly symmetric
    junctions (i.e. $r_\alpha=r_\beta$, $s_\alpha=s_\beta=s=1$), the current exactly vanishes.
     Thus $s_{++}$ Josephson junctions exhibit greater critical current values ($I_{c}(T \rightarrow 0)
      \sim e \Delta_{\beta}/\hbar$) compared to those expected for
      the $s_{\pm}$ case ($I_{c}(T \rightarrow 0) \sim e \Delta_{\alpha}/\hbar$).
       On the other hand, the peculiar functional form of the current-phase relation of
       the $s_{++}$ symmetry (panel (b)) indicates a certain fragility of this junction
        against bias current fluctuations. Indeed, a small bias fluctuation (less than
        few percent of the critical current) can cause, at low bias, a relevant phase fluctuation.
         This stochastic phase jump may represent a relevant source of voltage fluctuation
         across the junction which can also drive the system towards an ohmic regime.\\
Figures \ref{fig:phase_current}(c) and \ref{fig:phase_current}(d) are obtained by setting
$r_\alpha=1.8$ and $r_\beta=1$, while maintaining the remaining parameters at the same
values fixed in the upper panels of Figure \ref{fig:phase_current}. In particular, panel
(c) shows an $s_{\pm}$ current-phase relation that, as also evident from the Andreev levels
 (Figure \ref{fig:levels} (c)), undergoes a $\pi$-shift as the $s_\alpha$  parameter
 varies between the values $0.3$ and $1$. The $\pi$-shift is absent in panel (d) where
 the current-phase curve is shown for the $s_{++}$ symmetry. A $\pi$-junction presents
 a free energy $F(\varphi)=\Phi_0/2\pi \int_0^\varphi d \theta I(\theta)$ minimum at
  $\varphi=\pi$ rather than at $\varphi=0$. The existence of $\pi$-shifted junctions
  can be experimentally proved employing SQUID's measurements.\\
While the current-phase relation cannot be simply measured, the measurement of the
temperature dependence of the junction critical current is relatively easier. A
comparison with the experiments of this dependence represents very often an important
 tool contributing to identifying the nature of the superconducting pairing. Thus, in
  order to compare our results with the experimental findings, we have derived the
  temperature dependence of the maximum Josephson current (critical current). Figure
  \ref{fig:IcvsT} shows the temperature dependence of the critical current calculated
   for the $s_{\pm}$- ((a) and (c)) and $s_{++}$-symmetry ((b) and (d)) by setting
   the model parameters as done in Figs. \ref{fig:levels} and \ref{fig:phase_current}.
    Negative critical current values, like those presented in panel (c), indicates a
     $\pi$-shifted Josephson junction. A common feature of all the panels in Figure
     \ref{fig:IcvsT} is the violation of the Ambegaokar-Baratoff relation, which is
      usually reported in multiband S/I/S Josephson junctions \cite{ambegaokar_baratoff-violation}.
      The entity of such violation depends on the junction parameters and is more pronounced
       for the $s_{\pm}$ symmetry (see panels (a) and (c)), where the critical current \textit{vs}
        temperature curves develop a peculiar positive curvature. Panels (b) and (d)
        are described by a deformed Ambegaokar-Baratoff relation and suggest that a $\pi$
         junction cannot be observed for the $s_{++}$ symmetry. In this respect we observe
          that the $0-\pi$ transition reported in the present work (see Figure
           \ref{fig:phase_current}(c) and \ref{fig:IcvsT}(c)) is the result
           of a different mechanism compared to the one described in Ref. \cite{nap}.
            There the transition originates from the competition of two Andreev bound
            states carrying opposite current. As the temperature is lowered the state
             with higher energy eigenvalue is progressively excluded from the transport
             and a sign reversal of the critical current is observed. Here we predict
             the presence of a single electron-like bound state and thus a temperature
             activated $0-\pi$ transition cannot be observed as an intrinsic effect.
             However, direct tunneling ($\alpha \rightarrow \alpha$ or $\beta \rightarrow \beta$)
             and crossed tunneling ($\alpha \rightarrow \beta$ or $\beta \rightarrow \alpha$)
             effects provide supercurrent contribution with opposite sign and thus, depending
             on the relative strength of these contributions, a $\pi$-junction can be formed
             only assuming the $s_{\pm}$ symmetry. The relative strength of the direct and
             crossed tunneling contributions is controlled in the model by the overlap parameters
             $s_\alpha$, $s_\beta$, and $s$ which are characteristic of the interface. According
             to these arguments, a $0-\pi$ transition can occur by varying at least one overlap
             parameter. This can be done in different ways: (i) changing the temperature can induce
              a lattice deformation at the interface which is relevant in determining the overlap
               parameters; (ii) experiments performed under controllable pressure or strain allow
               to control lattice distortion and thus the overlap parameters. Both the methods
               (i) and (ii) can be used to induce the $0-\pi$ transition in multiband Josephson junctions.
The mechanism of formation of a $\pi$-junction described above can be also recovered in the
framework of a semiclassical theory of the $s_{\pm}/I/s_{\pm}$ Josephson junction as developed in Ref. \cite{deluca2013}.\\
\begin{figure}[htbp]
\centering
\includegraphics[width=10pc]{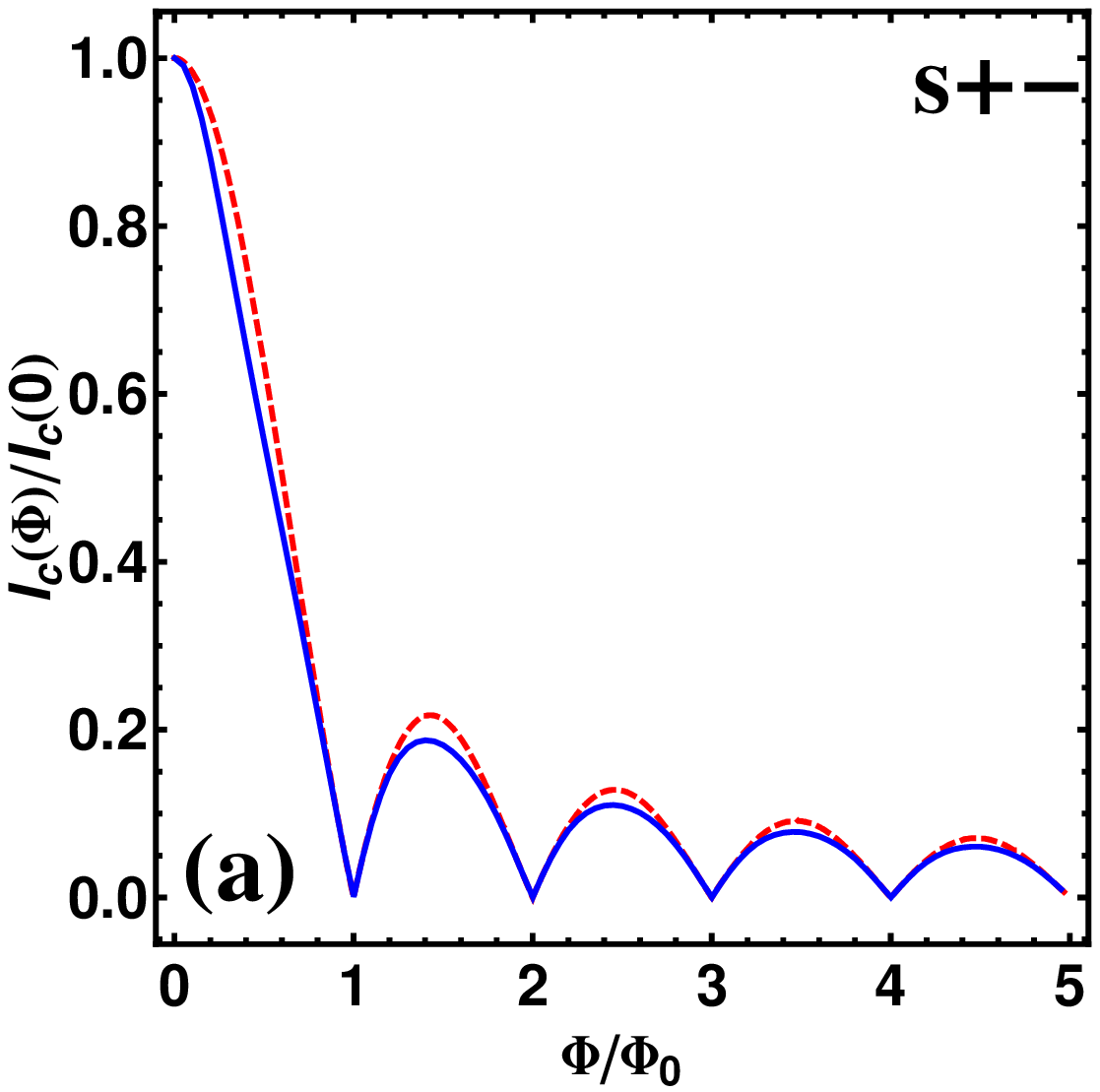}
\includegraphics[width=10pc]{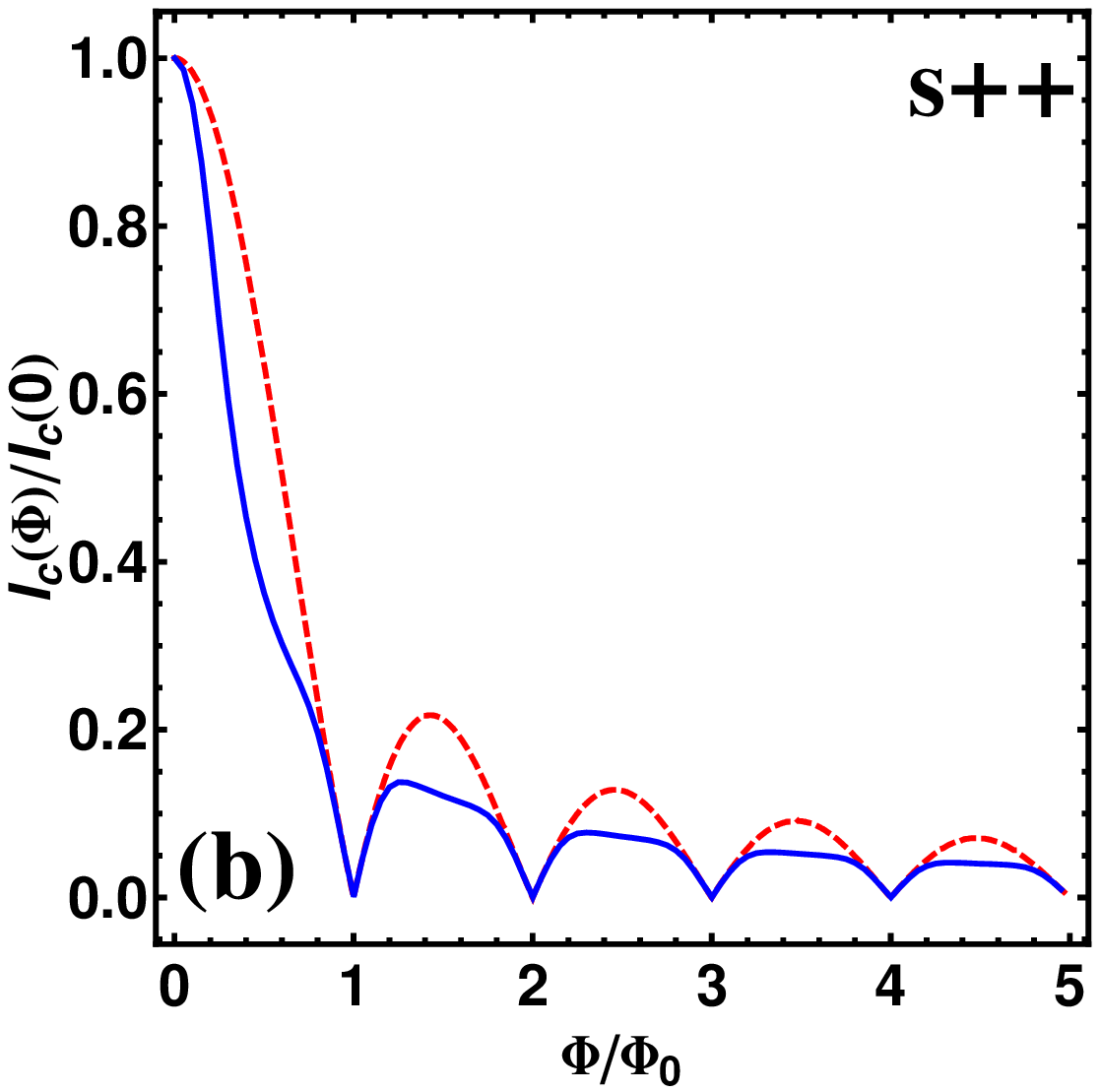}
 \caption{ Normalized critical current $I_{c}(\Phi)/I_{c}(0)$ as a function
  of the applied magnetic flux $\Phi/\Phi_{0}$ computed for the $s_{\pm}$
  (panel (a)) and the $s_{++}$ symmetry (panel (b)). Parameters are the same
   of the corresponding panels in Fig.~3: $Z=0.03$, $\gamma=3$, $r_{1}=1.8$,
    $r_{2}=1$, $s_{\alpha}=s_{\beta}=s=1$, $\chi_0=0.01$ and $T=0.001 T_{c}$.
     The dashed curve in each panel represents the Fraunhofer pattern for
      comparison.}\label{fig:fp}
\end{figure}
However, as clearly evidenced in Figs.~\ref{fig:phase_current}(a)-(b),
the $0-\pi$ transition is strongly affected by the bands effective mass
and can remain unobserved also for an $s_{\pm}$ junction. Under this
condition, the magnetic response of the system (see Appendix \ref{appendix:Ic-mfield}
 for details) provides an indirect probe of the harmonic content of the
 current-phase relation and can help in discriminating the pairing symmetry
 of the junction. In particular, the zero-field critical current
  $I_{c}(0)$ of the $s_{++}$ case (see Fig.~\ref{fig:phase_current}(b))
   is greater than the corresponding value obtained for the
   $s_{\pm}$ case (see Fig.~\ref{fig:phase_current}(a)).
   As a consequence, one expects that the harmonic content
    of the current-phase relation of the $s_{++}$ case
     strongly affects the low-field magnetic diffraction
      pattern of the junction. On the other hand, for
       the $s_{\pm}$ case, the high harmonic contribution
       is expected to be less important due to the lower
        value of zero-field critical current. The above
         arguments are confirmed by Fig.~\ref{fig:fp}
         where the normalized critical current $I_{c}(\Phi)/I_{c}(0)$
         is studied as a function of the applied magnetic flux $\Phi/\Phi_{0}$
          (normalized to the flux quantum $\Phi_{0}$) for the $s_{\pm}$
          (panel (a)) and the $s_{++}$ (panel (b)) symmetry, while the model
           parameters are fixed as done for the corresponding panels
           in Fig.~3. Interestingly, the $s_{++}$ Josephson junction
           evidences a critical current halving already for flux values
            of $\sim 0.4 \Phi_{0}$, while the critical current of the
             $s_{\pm}$ case is less sensitive to magnetic field effects.

\section{Conclusions}
\label{sec:conclusions}
We have formulated a minimal model of the dc Josephson effect
for multiband superconductors based on the quantum waveguide approach.
 The method is based on the analogy between multiband superconductors
 and multibranch networks recently suggested in Ref.~\cite{Araujo}.
 Accordingly, the subgap bound states wave functions (electron- and hole-like),
 relevant in describing the quantum transport within the short junction limit,
 are decomposed using the eigenstates of local branch Hamiltonians and the
  coefficients of such a decomposition are found imposing generalized
  boundary conditions on the wave functions. The boundary conditions, never
   used before for this problem, are direct generalization of those used
   in Ref.~\cite{Romeo} and incorporate a local scattering potential at the
    interface ($Z$ parameter) and band overlap factors $s_\alpha$, $s_\beta$,
     and $s$ which define the weight of each band in the quasiparticles transport.
      This provides an effective parametrization of the interface effects
      allowing to describe Josephson junctions ranging from the metallic ($Z\ll 1$)
       to the tunnel limit ($Z\gg 1$). The bulk properties of the superconducting
        bands are introduced using different effective masses ($r_{j}$ parameters)
         which can be determined by complementary experiments.
We have solved the scattering problem and we have determined the Andreev
 bound states spectrum and the normalized eigenfunctions. The Josephson
 current flowing through the junction has been computed using a second
 quantization approach which correctly reproduces results obtained using
  the phase derivative of the Andreev bound states spectrum formula.
  The second quantization method used in the derivation of the Josephson
   current generalizes the result presented in Ref.~\cite{furusaki}
   for a single band superconductor to the multiband case and represents
    one of the main results of this work. To provide a specific and
    relevant application of the theory, we have focused our treatment
     on FeBS-FeBS tunnel junctions. In particular, we have derived the
      current-phase relations and the critical currents of symmetric FeBS
      junctions modeled as superconducting systems with two relevant bands.
       We have analyzed $s_{++}$ and $s_{\pm}$ symmetries of the FeBS and
       we have shown that a $\pi$-junction can be observed only for $s_{\pm}$
       symmetry and under appropriate interface conditions, which are
       carefully discussed in the text. Further peculiar aspects of FeBS-FeBS
       tunnel junctions are also discussed. The above findings are relevant
       for the Josephson effect theory in multiband systems and can contribute
        to the current debate about the order parameter symmetry of
        iron-based materials.

\begin{acknowledgments}
 This work has been partially supported by the financial
 contribution of EU NMP.2011.2.2-6 IRONSEA project nr. 283141.
  The authors grateful acknowledge Roberto De Luca for useful
  discussions on the topic of this work.
\end{acknowledgments}

\appendix
\section{Determination of the scattering coefficients}
\label{appendix:matrix}
The matching conditions Eqs.~(\ref{eq:matching1})-(\ref{eq:matching2})
 provide the linear homogeneous system $ \mathbf{M} \cdot \mathbf{X}=0$,
  where $\mathbf{X}=(a_\alpha,..., d_\alpha, a_\beta,..., d_\beta)^{t}$ and $\mathbf{M}$
  is the following matrix
\begin{widetext}
\begin{equation}
\left(
  \begin{array}{cccccccc}
    s_\alpha v_\alpha & s_\alpha u_\alpha & -u_\alpha & -v_\alpha & 0 & 0 & 0 & 0 \\
    s_\alpha u_\alpha & s_\alpha v_\alpha & -e^{-i \varphi}v_\alpha & -e^{-i \varphi}u_\alpha & 0 & 0 & 0 & 0 \\
    0 & 0 & 0 & 0 &s_\beta v_\beta & s_\beta u_\beta & -u_\beta & -v_\beta \\
    0 & 0 & 0 & 0 & e^{-i\delta}s_\beta u_\beta & e^{-i\delta}s_\beta v_\beta
    & -e^{-i (\delta+\varphi)}v_\beta & -e^{-i (\delta+\varphi)}u_\beta \\
    s v_\alpha & s u_\alpha & 0 & 0 & 0 & 0 & -u_\beta & -v_\beta \\
    s u_\alpha & s v_\alpha & 0 & 0 & 0 & 0 & -e^{-i (\delta+\varphi)}v_\beta
     & -e^{-i (\delta+\varphi)}u_\beta \\
    -\frac{ v_\alpha (i + 2 r_\alpha Z)}{r_\alpha} &  \frac{ u_\alpha (i - 2 r_\alpha Z)}{r_\alpha} & \frac{i s_\alpha u_\alpha}{r_\alpha} &
     -\frac{i
     s_\alpha v_\alpha}{r_\alpha} & -\frac{i s v_\beta}{r_\beta} &
       \frac{i s u_\beta}{r_\beta} & \frac{i s_\beta u_\beta}{r_\beta} & -\frac{ i s_\beta v_\beta}{r_\beta} \\
    -\frac{ u_\alpha (i + 2 r_\alpha Z)}{r_\alpha} &  \frac{ v_\alpha
     (i - 2 r_\alpha Z)}{r_\alpha}  & \frac{i e^{- i\varphi} s_\alpha v_\alpha}{r_\alpha} & -\frac{i e^{- i\varphi}
    s_\alpha u_\alpha}{r_\alpha} & -\frac{i e^{- i\delta}
    s u_\beta}{r_\beta} &  \frac{i e^{- i\delta}
    s v_\beta}{r_\beta} & \frac{i e^{-i (\delta+\varphi)} s_\beta v_\beta}{r_\beta}
     & -\frac{i e^{-i (\delta+\varphi)} s_\beta u_\beta}{r_\beta}  \\
  \end{array}
\right).
\end{equation}
\end{widetext}
The homogeneous system admits non-trivial solution provided that
 the condition $det(\mathbf{M})=0$ (Eq.~(\ref{eq:spectrum}) of
 the main text) is fulfilled. Since $\mathbf{M}$ is not a maximal
  rank matrix, the normalization condition of the bound state
   wave function provides a further condition to determine the
    coefficients vector $\mathbf{X}$.

\section{Magnetic field dependence of the critical current}
\label{appendix:Ic-mfield}
We describe the response of a Josephson junction under the
 application of a sufficiently weak external magnetic field
  $\vec{H}=H \hat{z}$ parallel to the $z$-axis. We explicitly
   assume the small junction limit in which the transverse
   dimension $L$ of the junction (parallel to the $y$-axis)
    is comparable or smaller than the Josephson penetration
    length $\lambda_j$ ($L\lesssim \lambda_j$) \cite{barone-book}.
     The junction region, located at $x=0$, presents a reduced
      pairing potential and thus experiences the maximum
      magnetic field value, while inside the electrodes
       the polarizing effect of the field is effectively
       screened by supercurrents. Due to this, the spatial
        dependence of the magnetic field is given by
         $\vec{H}=H \hat{z} \exp(-|x|/\lambda)$. Since
          we are interested in the bulk effect of
          $\vec{H}$, observing that $\int_{-\infty}^{\infty}\vec{H}
           \cdot \hat{z}dx=2\lambda H$, we can approximate the
            spatial dependence of the field according to the
            expression $\vec{H}=\hat{z} \frac{\Phi}{L}\delta(x)$,
            where we introduced the magnetic flux $\Phi=2\lambda L H$
             induced by the external field and the Dirac delta
             function $\delta(x)$. The magnetic field $\vec{H}=
             \vec{\nabla} \times \vec{A}$ can be expressed in
             terms of the vector potential $\vec{A}=(-\frac{\Phi y}{L}\delta(x),0,0)$
             which is not affected by self-fields in the considered limit.
              Assuming that the Zeeman term is effectively screened in the
               bulk of the electrodes, the presence of $\vec{H}$ affects
               the BdG branch Hamiltonian $H_{BdG}^{(j)}$ (Eq.~\ref{eq:hamilton})
                only through the substitution $\hat{p}_{x}\rightarrow \hat{p}_{x}-e A_{x}$.
                The vector potential $A_{x}$ can be gauged away by means
                of the unitary transformation
\begin{equation}
U=\exp(i \pi \frac{\Phi}{\Phi_{0}}\frac{y}{L}\theta(x)\hat{\sigma}_{z}),
\end{equation}
with $\hat{\sigma}_{z}$ a Pauli matrix acting on the
 particle-hole space and $\Phi_{0}=h/(2|e|)$ the elementary
 flux quantum. The transformed branch Hamiltonian $\widetilde{H}^{(j)}_{BdG}=U^{\dag}H^{(j)}_{BdG}U$
 under the action of $U$ can be obtained with the following substitutions:
\begin{eqnarray}
&& U^{\ast}_{11}\hat{H}_{j}(x;A_{x})U_{11}\rightarrow -\frac{\hbar^2 \partial^{2}_{x}}{2m_{j}}\nonumber\\
&& -U^{\ast}_{22}\hat{H}^{\ast}_{j}(x;A_{x})U_{22}\rightarrow \frac{\hbar^2 \partial^{2}_{x}}{2m_{j}}\nonumber\\
&& \Delta_{j}(x)U^{\ast}_{11}U_{22}\rightarrow \widetilde{\Delta}_{j}(x),
\end{eqnarray}
while the BdG state associated to $\widetilde{H}^{(j)}_{BdG}$ is given by $\widetilde{\Psi}_{j}=U^{\dag}\Psi_{j}$.
The pair potential $\widetilde{\Delta}_{j}(x)$ takes the same mathematical
 structure of the one presented in Eq.~(\ref{eq:pair_potential}), the only
  action of the unitary transformation being the substitution of the phase
  difference $\varphi$ with $\widetilde{\varphi}=\varphi-2\pi \frac{\Phi}{\Phi_{0}}\frac{y}{L}$.
   As a consequence the phase difference between the two sides of the junction is
   modulated along the junction and this modulation is the source of the magnetic
   diffraction pattern affecting the critical current of the junction. Under our
   assumptions, the $y$-dependence of the pair potential is adiabatic compared to
    the microscopic scale of the problem. This statement can be rigorously proved
    using the two-scale perturbation theory \cite{strogatz}. In particular the
     wave vector, $q=\frac{2\pi}{L}\frac{\Phi}{\Phi_{0}}$, modulating the
      superconducting phase is much smaller than the Fermi wave vectors
      $r_{j}k_F$ and thus the phase modulation along the $y$-direction enters
       only parametrically in the one dimensional problem described in the main text.
       The validity of these arguments also requires that the Cooper pairs tunneling
        with normal incidence represents the dominant microscopic process and thus
        the energy associated with transverse modes is negligible compared to the
        Fermi energy. When the energy of the transverse modes becomes relevant
        (e.g. for very short junctions with $L\ll \lambda $) on the Fermi energy
        scale, a full treatment of the transverse degrees of freedom is required.
Hereafter, we focus our attention on the case of adiabatic phase variation of $\widetilde{\varphi}$
along the transverse dimension of the junction. Under this assumption, once the
current-phase relation $I_{J}(\varphi)$ has been obtained according to the procedure
given in the main text, the magnetic field dependence of the critical current of the
junction is obtained according to the formula:
\begin{equation}
\label{eq:app-icvsfield}
I_{c}(\Phi)=\max_{\varphi \in [0,2\pi]}\int_{-L/2}^{L/2}\frac{d y}{L}I_{J}\left(\varphi-2\pi \frac{\Phi}{\Phi_{0}}\frac{y}{L}\right),
\end{equation}
where the current-phase relation presents a parametric dependence on $y$.
A further progress can be done observing that the current-phase relation
is an odd function of the phase difference and thus it can be written as
$I_{J}(\varphi)=\sum_{n=0}^{\infty} a_{n}\sin(n\varphi)$. As a consequence,
 Eq.~(\ref{eq:app-icvsfield}) takes the form:
\begin{equation}
\label{eq:app-icvsfield2}
I_{c}(\Phi)=\max_{\varphi \in [0,2\pi]}\sum_{n=0}^{\infty} a_{n}\sin(n\varphi)
 \frac{\sin(n\frac{\pi \Phi}{\Phi_{0}})}{n\frac{\pi \Phi}{\Phi_{0}}},
\end{equation}
where the coefficients $a_{n}$ can be directly extracted by $I_{J}(\varphi)$
using the relation:
\begin{equation}
a_{n}=\frac{1}{\pi}\int_{0}^{2\pi}I_{J}(\varphi)\sin(n \varphi) d\varphi,
\end{equation}
where we explicitly used the orthogonality condition $\int_{0}^{2\pi}\sin(n \varphi)\sin(m \varphi)
d\varphi=\pi \delta_{n,m}$. Equation~(\ref{eq:app-icvsfield2}) provides the Fraunhofer diffraction pattern,
\begin{equation}
I_{c}(\Phi)/I_{c}(0)=\Bigl|\frac{\sin(\frac{\pi \Phi}{\Phi_{0}})}{\frac{\pi \Phi}{\Phi_{0}}}\Bigl|\nonumber,
\end{equation}
when a single-harmonic current-phase dependence $I_{J}(\varphi)=a_{1} \sin(\varphi)$ is
 considered, while deviations are expected if the high-harmonic contribution is not negligible.
  For the above reasons, the magnetic diffraction pattern $I_{c}(\Phi)$ is an indirect probe
  of the harmonic content of the current-phase relation of the junction. Our treatment of the
  magnetic field effects on the junction justifies the validity of the approach proposed
   in Refs.~\cite{goldobin,sperstad} and the classical argument given in Ref.~\cite{barone-book}.

\newpage 


\begin{thebibliography}{apssamp}
\bibitem{multiband-superc1959}H. Suhl, B. T. Matthias, and L. R. Walker, Phys.
Rev. Lett. \textbf{3}, 552 (1959); V. A. Moskalenko, Fiz. Met. Metalloved. \textbf{8}, 503 (1959).
\bibitem{bcs}J. Bardeen, L. N. Cooper, J. R. Schrieffer, Phys. Rev. \textbf{108}, 1175 (1957).
\bibitem{mgb2}J. Nagamatsu, N. Nakagawa, T. Muranaka, Y. Zenitani, and J. Akimitsu,
 Nature (London) \textbf{410}, 63 (2001).
\bibitem{kami}Y. Kamihara, T. Watanabe, M. Hirano and H. Hosono, J. Am.
Chem. Soc. {\bf 130}, 3296 (2008).
\bibitem{stew}G. R. Stewart, Rev. Mod. Phys. {\bf 83}, 1589 (2011).
\bibitem{giubileo2001}F. Giubileo, D. Roditchev, W. Sacks, R. Lamy, D. X.
 Thanh, J. Klein, S. Miraglia, D. Fruchart, J. Marcus, Ph. Monod, Phys. Rev. Lett. \textbf{87}, 177008 (2001).
\bibitem{s_pm_symm} I. I. Mazin, D. J. Singh, M. D. Johannes, and M. H.
 Du, Phys. Rev. Lett. \textbf{101}, 057003 (2008); K. Kuroki, S. Onari, R. Arita, H. Usui, Y. Tanaka,
H. Kontani, H. Aoki, Phys. Rev. Lett. \textbf{101}, 087004 (2008).
\bibitem{gonn}R. S. Gonnelli, D. Daghero, G. A. Ummarino, A. Calzolari,
M. Tortello, V. A. Stepanov, N. D. Zhigadlo, K. Rogacki, J.
Karpinski, F. Bernardini, and S. Massidda, Phys. Rev. Lett. {\bf 97},
037001 (2006).
\bibitem{dagh} D. Daghero, M. Tortello, G. A. Ummarino and R. S. Gonnelli,
 Rep. Prog. Phys. {\bf 74}, 124509 (2011).
\bibitem{park1}W. K. Park, J. L. Sarrao, J. D. Thompson, and L. H. Greene,
Phys. Rev. Lett. {\bf 100}, 177001 (2008).
\bibitem{kash}S. Kashiwaya, Y. Tanaka, M. Koyanagi, H. Takashima, and
K. Kajimura, Phys. Rev. B {\bf 51}, 1350 (1995).
\bibitem{chen}T. Y. Chen, Z. Tesanovic, R. H. Liu, X. H. Chen, and C. L.
Chien, Nature (London) {\bf 453}, 1224 (2008).
\bibitem{lu}X. Lu, W. K. Park, H. Q. Yuan, G. F. Chen, G. L. Luo, N. L. Wang,
A. S. Sefat, M. A. McGuire, R. Jin, B. C. Sales, D. Mandrus,
J. Gillett, S. E. Sebastian, and L. H. Greene, Supercond. Sci.
Technol. {\bf 23}, 054009 (2010).
 \bibitem{ili}E. Il'ichev, M. Grajcar, R. Hlubina, R. P. J. IJsselsteijn,
  H. E. Hoenig, H.-G. Meyer, A. Golubov, M. H. S. Amin, A. M. Zagoskin,
  A. N. Omelyanchouk, and M. Yu. Kupriyanov, Phys. Rev. Lett. {\bf 86}, 5369 (2001).
\bibitem{testa1} G. Testa, A. Monaco, E. Esposito, E. Sarnelli, D.-J. Kang,
S. H. Mennema, E. J. Tarte and M. G. Blamire, Appl.
Phys. Lett. {\bf 85}, 1202 (2004).
\bibitem{testa2} G. Testa, E. Sarnelli, A. Monaco, E. Esposito, M. Ejrnaes,
D.-J. Kang, E. H. Mennema, E. J. Tarte, and M. G. Blamire,
Phys. Rev. B {\bf 71}, 134520 (2005).
\bibitem{d0d0}E. Sarnelli, M. Adamo, S. De Nicola, S. Cibella, R. Leoni and
C. Nappi, Supercond. Sci. Technol. {\bf 26} 105013 (2013).
\bibitem{Araujo} M. A. N. Araujo, P. D. Sacramento, Phys. Rev. B {\bf 79}, 174529 (2009).
 \bibitem{Xia} J.-B. Xia, Phys. Rev. B {\bf 45}, 3593 (1992).
 \bibitem{Romeo} F. Romeo, R. Citro, Phys. Rev. B {\bf 91}, 035427 (2015).
 \bibitem{wu} C. H. Wu, W. C. Chang, J. T. Jeng, M. J. Wang, Y. S. Li, H. H. Chang,
and M. K. Wu, Appl. Phys. Lett. {\bf 102}, 222602 (2013).
 \bibitem{burm2}A. V. Burmistrova and I. A. Devyatov, Europhys. Lett. {\bf 107}, 67006 (2014).
 \bibitem{burm1}A. V. Burmistrova, I. A. Devyatov, A. A. Golubov, K. Yada, and Y. Tanaka,
 J. Phys. Soc. Jpn. {\bf 82}, 034716 (2013).
 \bibitem{golu} A. A. Golubov, I. I. Mazin, Appl. Phys. Lett. {\bf 102}, 032601 (2013).
  \bibitem{nap} C. Nappi, S. De Nicola, M. Adamo, E. Sarnelli, Europhys. Lett. {\bf 102} 47007(2013).
 \bibitem{moor} A. Moor, A. F. Volkov, K. B. Efetov, Phys. Rev. B  {\bf 87}, 100504 (2013).
\bibitem{da}W. Da, L. Houng-Yan and W. Qiang-Hua,  Chin. Phys. Lett. {\bf 30}, 077404 (2013).
\bibitem{stan}V. G. Stanev, A. E. Koshelev, Phys. Rev. B {\bf 86}, 174515 (2012).
 \bibitem{lin}S. Z. Lin, Phys. Rev. B {\bf 86}, 014510 (2012).
\bibitem{apos}S. Apostolov, A. Levchenko, Phys. Rev. B {\bf 86}, 224501 (2012).
\bibitem{kosh} A. E. Koshelev, V. G. Stanev, Europhys. Lett. {\bf 96}, 27014 (2011).
\bibitem{berg} E. Berg, Phys. Rev. Lett. {\bf 106}, 147003 (2011).
\bibitem{ota}Y. Ota, N. Nakai, H. Nakamura, M. Machida, D. Inotani, Y. Ohashi,
T. Koyama, and H. Matsumoto, Phys. Rev. B {\bf 81}, 214511 (2010).
 \bibitem{erin}Yu. Erin, A. N. Omel'yanchuk, Low. Temp. Phys. {\bf 36} 969 (2010).
\bibitem{park} D. Parker, I. I. Mazin, Phys. Rev. Lett. {\bf 102}, 227007 (2009).
\bibitem{wu1}J. Wu, P. Phillips, Phys. Rev. B {\bf 79}, 092502 (2009).
\bibitem{ng} T. K. Ng and N. Nagaosa, Europhys. Lett. {\bf 87}, 17003 (2009).
\bibitem{golu2} A. A. Golubov, A. Brinkman, Y. Tanaka, I. I. Mazin, and O. V.
Dolgov, Phys. Rev. Lett. {\bf 103}, 077003 (2009).
 \bibitem{lind}J. Linder, I. B. Sperstad, A. Sudb{\o}, Phys. Rev. {\bf 80}, R020503 (2009).
\bibitem{tsai} Wei-Feng Tsai, Dao-Xin Yao, B. Andrei Bernevig, and JiangPing Hu, Phys. Rev. B {\bf 80}, 012511 (2009).
\bibitem{chen2} Wei-Qiang Chen, Fengjie Ma, Zhong-Yi Lu, and Fu-Chun Zhang, Phys. Rev. Lett. {\bf 103}, 207001 (2009).
\bibitem{seba} S. D\"{o}ring, S. Schmidt, D. Reifert, M. Feltz, M. Monecke,
N. Hasan, V. Tympel, F. Schmidl, J. Engelmann, F. Kurth,
K. Iida, I. M\"{o}nch, B. Holzapfel, P. Seidel, J. Supercond. Nov. Magn. {\bf 28}, 1117 (2015).
\bibitem{dor} S. D\"{o}ring, S. Schmidt, F. Schmidl, V. Tympel, S. Haindl, F. Kurth,
K. Iida, I. M\"{o}nch, B. Holzapfel, and P. Seidel, Superc. Sci. Technol. {\bf 25}, 084020 (2012).
\bibitem{dor2} S. D\"{o}ring, M. Monecke, S. Schmidt, F. Schmidl,
 V. Tympel, J. Engelmann, F. Kurth, K. Iida, S. Haindl, I. M\"{o}nch,
B. Holzapfel, and P. Seidel, J. App. Phys. {\bf 115}, 083901 (2014).
\bibitem{schm}S. Schmidt, S. D\"{o}ring, F. Schmidl, V. Tympel, S. Haindl,
K. Iida, F. Kurth, B. Holzapfel, and P. Seidel, IEEE Trans. Appl. Supercond. {\bf 23}, 7300104 (2013).
\bibitem{lee}S. Lee, J. Jiang, J. D. Weiss, C. M. Folkman, C. W. Bark, C. Tarantini, A. Xu,
D. Abraimov, A. Polyanskii, C. T. Nelson, Y. Zhang, S. H. Baek, H. W. Jang,
A. Yamamoto, F. Kametani, X. Q. Pan, E. E. Hellstrom, A. Gurevich, C. B. Eom,
and D. C. Larbalestier, Appl. Phys. Lett. {\bf 95}, 212505 (2009).
\bibitem{kata1} T. Katase, Y. Ishimaru, A. Tsukamoto, H. Hiramatsu, T. Kamiya, K.
Tanabe, and H. Hosono, Appl. Phys. Lett. {\bf 96}, 142507 (2010).
\bibitem{kata2}T. Katase, Y. Ishimaru, A. Tsukamoto, H. Hiramatsu, T. Kamiya, K.
Tanabe, and H. Hosono, Supercond. Sci. Technol. {\bf 23}, 082001 (2010).
\bibitem{sarn} E. Sarnelli, M. Adamo, C. Nappi, V. Braccini, S. Kawale, E. Bellingeri, C. Ferdeghini, Appl. Phys. Lett.
{\bf 104}, 162601 (2014).
\bibitem{wu2}C. H. Wu, W. C. Chang, J. T. Jeng, M. J. Wang, Y. S. Li, H. H. Chang,
and M. K. Wu, Appl. Phys. Lett. {\bf 102}, 222602 (2013).
\bibitem{baro} C. Barone, F. Romeo, S. Pagano, M. Adamo, C. Nappi, E. Sarnelli, F. Kurth, K. Iida,
 Scientific Reports {\bf 4}, 6163 (2014).
\bibitem{seidel} P. Seidel, Superc. Sci. Technol. {\bf 24}, 043001 (2011).
\bibitem{blond} G. E. Blonder, M. Thinkham and T. M. Klapwijk, Phys. Rev. B {\bf 25} 4515 (1982).
\bibitem{nota-Zj} Observing that $m r_{j}^{2}=m_{j}$ and $r_{j}k_{F}=k_{F}^{(j)}$, the quantity
 $r_{j}Z=\frac{mU_{0}}{\hbar^{2}k_{F}}r_{j}$ can be easily indentified as the
  Blonder-Thinkham-Klapwijk parameter describing the scattering of a quasiparticle
  of mass $m_{j}$ and wavevector $k_{F}^{(j)}$, i.e. $Z_{j}=\frac{m_{j}U_{0}}{\hbar^{2}k_{F}^{(j)}}$.
\bibitem{benakk} C. W. J. Beenakker, Phys. Rev. Lett. {\bf 67}, 3836 (1991).
\bibitem{nota_diff_op} The first quantization operators $\overleftarrow{\partial}_{x}$ or
$\overrightarrow{\partial}_{x}$ act over the second quantization fields $\hat{A}$ and $\hat{B}$ as follows:\\
     $\hat{A}[\alpha \overrightarrow{\partial}_{x}+\beta \overleftarrow{\partial}_{x}]\hat{B}=
     \alpha \hat{A}\ (\partial_{x}\hat{B})+\beta (\partial_{x}\hat{A}) \ \hat{B}$, with $\alpha$, $\beta$ c-numbers.
\bibitem{nota_current} In the absence of magnetic potentials, the intensity of the charge current
 $\bar{J}_{ch, \sigma}=\sum_{j}\langle\hat{\psi}^{\dag}_{j \sigma} \check{\mathcal{J}}_{j}
 \hat{\psi}_{j \sigma}\rangle$ sustained by spin $\sigma \in \{\uparrow,\downarrow\}$ electrons
 does not depend on the spin projection $\sigma$, i.e. $\bar{J}_{ch, \uparrow}=\bar{J}_{ch,
  \downarrow}$. As a consequence $\bar{J}_{ch}$ can be computed according to the
   formula $\bar{J}_{ch}=2\bar{J}_{ch, \uparrow}$, as performed in the main text.
\bibitem{nota_current2} The charge current $\bar{J}_{ch}$ can be expressed
in units of $|e|\Delta_{\alpha}(0)/\hbar$ according to the formula:
 $\mathcal{I}=2 \cdot \chi_{0}^{-1}\sum_{j}r_{j}^{-1}u_{j}v_{j}k_{F}^{-1}(|c_{j}|^2-|d_{j}|^2)\tanh[E_B/(2 k_{B}T)]$,
  with $\bar{J}_{ch}=\frac{|e|\Delta_{\alpha}(0)}{\hbar}\mathcal{I}$.

\bibitem{furusaki} A. Furusaki, Superlattice and Microstructures {\bf 25}, 809 (1999).

\bibitem{ambegaokar_baratoff-violation} A. Brinkman, A. A. Golubov, H.
Rogalla, O. V. Dolgov, J. Kortus, Y. Kong, O. Jepsen, O. K. Andersen,
Phys. Rev. B \textbf{65}, 180517(R) (2002); A. Brinkman, A. A. Golubov,
 M. Yu. Kupriyanov, Phys. Rev. B \textbf{69}, 214407 (2004); Ke Chen,
  C. G. Zhuang, Qi Li, Y. Zhu, P. M. Voyles, X. Weng, J. M. Redwing,
R. K. Singh, A. W. Kleinsasser, X. X. Xi, Appl. Phys. Lett. \textbf{96},
 042506 (2010).
\bibitem{deluca2013} R. De Luca,
 Eur. Phys. J. B \textbf{86}, 294 (2013); see Equation 17.
\bibitem{strogatz}S. H. Strogatz,
 \textit{Nonlinear Dynamics and Chaos: With Applications
 to Physics, Biology, Chemistry, and Engineering}. (Perseus Books, Cambridge, 1994); see Chap.7.
\bibitem{goldobin} E. Goldobin, D. Koelle, R. Kleiner,
 and A. Buzdin, Phys. Rev. B \textbf{76}, 224523 (2007).
\bibitem{sperstad}I. B. Sperstad, J. Linder,
and A. Sudb{\o}, Phys. Rev. B \textbf{80}, 144507 (2009).
\bibitem{barone-book}A. Barone and G. Patern\`{o},
\textit{ Physics and Applications of the Josephson effect}, (John Wiley \& Sons, New York, 1982).


\end{thebibliography}
\end{document}